# A new charge-transfer complex in UHV co-deposited tetramethoxypyrene and tetracyanoquinodimethane

K. Medjanik,[1] S. Perkert,[1] S. Naghavi,[2] M. Rudloff,[3] V. Solovyeva,[3] D. Chercka,[4] M. Huth,[3] S. A. Nepijko,[1] T. Methfessel,[1] C. Felser,[2] M. Baumgarten,[4] K. Müllen,[4] H.J. Elmers,[1] and G. Schönhense[1*]

[1] *Institut für Physik, Johannes Gutenberg-Universität, Staudingerweg 7, D-55128 Mainz, Germany*

[2] *Institut für Analytische und Anorganische Chemie, Johannes Gutenberg-Universität, Staudingerweg 9, D-55128 Mainz, Germany*

[3] *Physikalisches Institut, Goethe-Universität, Max-von-Laue-Str. 1, D-60438 Frankfurt am Main, Germany*

[4] *Max-Planck-Institut für Polymerforschung, Ackermannweg 10, D-55128 Mainz, Germany*

**Abstract**

UHV-deposited films of the mixed phase of tetramethoxypyrene and tetracyanoquinodimethane (TMP$_1$-TCNQ$_1$) on gold have been studied using ultraviolet photoelectron spectroscopy (UPS), X-ray-diffraction (XRD), infrared (IR) spectroscopy and scanning tunnelling spectroscopy (STS). The formation of an intermolecular charge-transfer (CT) compound is evident from the appearance of new reflexes in XRD ($d_1$ = 0.894 nm, $d_2$ = 0.677 nm). A softening of the CN stretching vibration (red-shift by 7 cm$^{-1}$) of TCNQ is visible in the IR spectra, being indicative of a CT of the order of 0.3 $e$ from TMP to TCNQ in the complex. Characteristic shifts of the electronic level positions occur in UPS and STS that are in reasonable agreement with the prediction of from DFT calculations (Gaussian03 with hybrid functional B3LYP). STS reveals a HOMO-LUMO gap of the CT complex of about 1.25 eV being much smaller than the gaps (>3.0 eV) of the pure moieties. The electron-injection and hole-injection barriers are 0.3 eV and 0.5 eV, respectively. Systematic differences in the positions of the HOMOs determined by UPS and STS are discussed in terms of the different information content of the two methods.

PACS numbers: 68.37.Ef, 71.15.Mb, 79.60.-i

# 1. Introduction

Novel charge-transfer (CT) compounds based on organic molecules with π-conjugated ring structures attract high attention because they offer the possibility to tailor the electronic structure of organic electronics devices.[1] The wavefunctions of the electronic π-systems of the donor and acceptor constituents can be sensitively tuned *via* a change of the charge density by adding specific ligand groups at the periphery of the molecule. Characteristic for such compounds is a highly correlated electronic ground state that can exhibit superconductivity, spin density waves, or charge density waves.[2] Thus, this class of compounds is highly interesting not only because of potential applications but also for fundamental research.

Typically, these donor acceptor mixtures are crystallized in solution where the formation of a CT complex can often be checked by its characteristic color.[3] Solution-growth of crystals, however, is prohibitive for the standard techniques of electronic structure investigations because of the unavoidable surface contaminations. Surface-sensitive spectroscopies like UPS and STS require atomically clean surfaces. Classical preparation methods like ion sputtering combined with heating fail because of the sensitivity of the organic materials on such treatments. Thus, in-situ co-deposition in ultrahigh vacuum (UHV) is the only way to obtain suitable samples for surface sensitive techniques. Recently, we have succeeded in the UHV-deposition of donor - acceptor mixed phases for the classical system TCNQ/BEDT-TTF[4] (bisethylenedithiotetrathiafulvalene) and for the novel donor - acceptor pair hexamethoxycoronene - coronene-hexaone,[5] both moieties being derived from the same parent molecule using a novel synthesis route.[6]

The synthetic approach developed facilitated the realization of new sets of donor and acceptor molecules with a systematic variation of the electronic properties *via* the type and number of substituents at the periphery. Retaining the aromatic ring system, the electron affinity and ionization energy can be controlled in a wide range from typical donor behavior for different degrees of methoxy substitution to strong acceptor behavior for the corresponding ketones. In the co-deposited films of the hexamethoxy- and hexaketo-derivatives of coronene the CT is relatively weak.[5] For the present investigation we selected the combination of the classical strong acceptor TCNQ with a novel donor molecule TMP (in 1:1 stoichiometry), derived from the parent molecule pyrene by substitution of four methoxy groups. Both molecules have similar sizes as sketched in Fig. 1a. In cyclovoltammetry (CV) of TMP is oxidized at 0.7 eV against ferrocene, which corresponds to an ionization energy of 5.5 eV with respect to vacuum. This indicates that TMP is a moderate electron donor. The strong acceptor TCNQ is reduced against ferrocene and the corresponding electron affinity level is estimated at 4.6 eV below the vacuum level.[7]

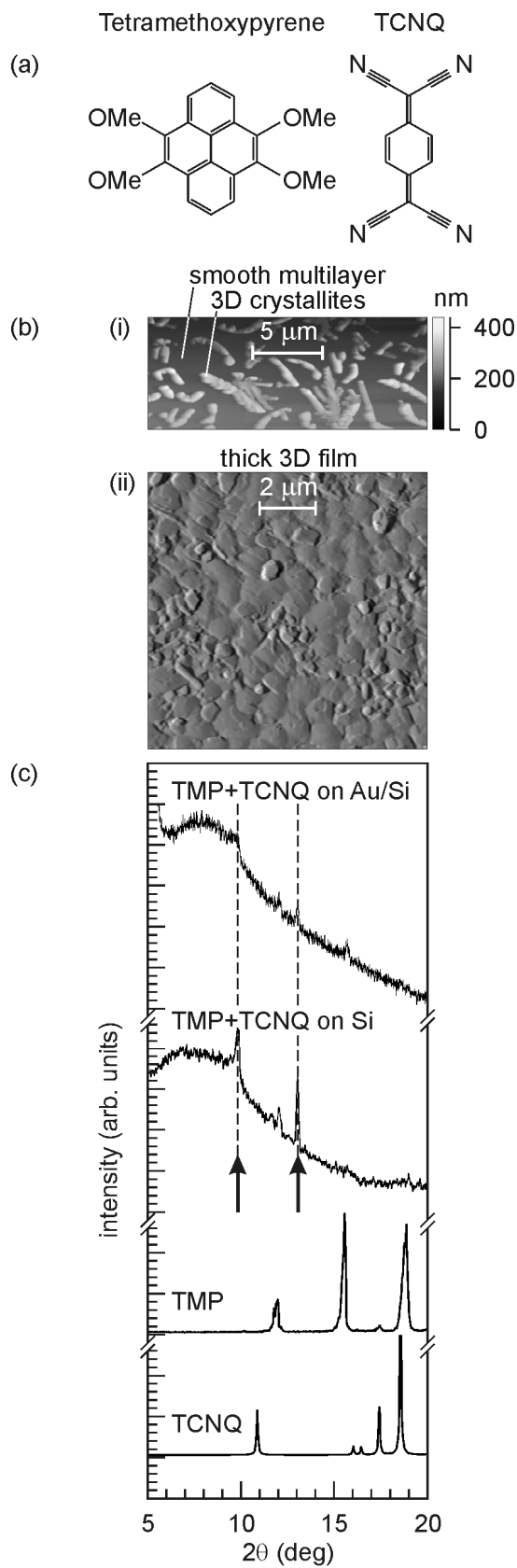

**FIG. 1.** Plot of the molecular structures (a), AFM images (b) and X-ray diffractograms (c) of thin films of the mixed phase $TCNQ_1$-$TMP_1$. Image (i) in (b) shows the coexistence of a smooth multilayer

covered with 3D crystallites. Image (ii), taken at higher coverage, shows a thick 3D film with crystallite sizes in the micrometer range. The X-ray diffraction patterns (c) for TMP-TCNQ films grown on Si and Au/Si substrates reveal new diffraction peaks (marked by arrows) in comparison with the reference diffraction patterns for pure TMP and TCNQ, see text.

The combination of TCNQ with an unsubstituted PAH has been studied earlier by Chi *et al.*[8] who found a CT of the order of 0.2-0.3 in a 1:1 stoichiometric crystal of TCNQ and coronene. This amount of CT was larger than observed for other hydrocarbon-based CT complexes in earlier work.[9-13] TCNQ-coronene showed semiconducting behavior with a moderate transport gap of 0.49 eV, along with a rather high charge carrier mobility of 0.3 cm$^2$/Vs.

Organic-metal interfaces have been investigated systematically, in particular for systems where CT occurs (see excellent reviews in Refs. [14-16]).

The workfunction is a very sensitive indicator of a CT between metal and organic layer. The compound has frontier energy levels that are hybrids of the respective frontier levels of the individual donor and acceptor molecule[17] as directly observed for the acceptor-donor pair F$_4$TCNQ - α-sexithiophene on a Au(111) surface.[18] Organic-metal interface states can show considerable energy dispersion because they result from a hybridization of metal and molecular states leading to anisotropic hybrid bands.[19]

The purpose of the present work was to investigate the electronic structure of the mixed phase of TCNQ and TMP by various spectroscopic techniques. The results are compared to those obtained from the respective pristine molecular films. The organic films were grown on Au surfaces by (co-)sublimation in UHV. The spectra give information on the interface of the organic films and the metallic substrate. Clear signatures of an interface CT (between metal and organic layer) and an inter-molecular CT in the compound mixture have been found. The formation of a novel CT phase is corroborated by the appearance of new reflexes in XRD and a red shift of the CN stretching vibration of TCNQ, observed in infrared spectroscopy. Density functional theory (DFT) calculations have been performed yielding the energy scheme and symmetries of valence orbitals of the pure moieties and of the compound. We find reasonable agreement between calculated and measured spectra. In co-deposited films the highest occupied molecular orbital (HOMO) derived from TMP and the lowest unoccupied molecular orbital (LUMO) derived from TCNQ are both shifted towards the Fermi level of the metal substrate. The HOMO-LUMO gap is reduced from 3.1 eV (3.16 eV) for pure TCNQ (TMP) to 1.25 eV in the CT complex as directly measured using STS.

## 2. UHV-Deposition of the TMP-TCNQ mixed phase

For UHV deposition of high-purity films we used an evaporator with graphite crucible in an oxygen-free high conductivity (OFHC) copper holder with thermocoax heating element and NiCr-CuNi thermoelement for temperature regulation. Using a differentially pumped load-lock system, the evaporator could be changed and refilled without breaking UHV. The crucible was placed at a distance of 40 mm from the sample in measuring position in front of the electron spectrometer. In this way, the evolution of the spectra could be observed in small coverage steps during film deposition. The TMP-TCNQ mixed phase can be deposited in very good quality by loading the crucible with a mixture of both molecules. The base pressure before evaporation was about $3 \cdot 10^{-10}$ mbar. Clean Au films were used as substrates with high electronegativity, providing a good reference for the Fermi energy. About 20 nm of Au were evaporated onto Si(100)-wafers immediately before deposition of the organic films. The workfunction of the fresh Au films was consistently 5.3 eV. This indicates that the textured Au surface exhibits Au(111) facets. The substrate was kept at room temperature, the crucible temperatures were 120° C, 135° C and 100° C for the compound, pure TMP and TCNQ, respectively.

Fig. 1a shows the structure of the molecules. Note that the methyl groups (Me = $CH_3$) are not planar, i.e. TMP does not posses a mirror plane. Fig. 1b shows AFM-images (atomic force microscopy) for two samples with different coverages. At lower coverage (sample (i)) 3D-crystallites with lateral sizes and height in the 100 nm range appear on a smooth background. The crystallites tend to form rows with lengths of up to several micrometers. At higher coverage (sample (ii)) the crystallites cover the whole surface. The films look dark orange giving evidence for an optical gap in the visible spectral range (films of pure TMP and TCNQ look yellow). In the photoelectron spectra corresponding to sample (i) the Fermi edge of the gold substrate was not visible. This means that the flat parts between the crystallites are not bare Au but are completely covered by the organic film. This suggests that the complex grows in a Stranski-Krastanov-like mode where an initial smooth multilayer covers the whole surface (smooth areas in image (ii)) before 3D crystallites are formed. We have observed similar microcrystals with sizes in the sub-micrometer range for the TCNQ/BEDT-TTF mixed phase.[4]

As a consequence of slightly different evaporation rates, the partial vapour pressures of the two moieties in the molecular beam change with time. Despite of this time-dependent deviation from the 1:1 mixture in the molecular beam, the stable compound favours a 1:1 composition of the mixed phase. We believe that this is due to small sticking coefficients of the molecules for a substrate held at room temperature. As a consequence, surplus donor or acceptor molecules, which are not subject to the CT phase formation, readily desorb from the surface. However, for sufficiently high vapour pressure pure TMP and/or pure TCNQ crystallites start to grow in coexistence with the mixed phase.

X-ray diffractometry was performed employing a Bruker D8 diffractometer with a Cu anode in parallel mode using a Goebel mirror. In Fig. 1c two diffraction patterns ($\theta - 2\theta$ scans) of TMP-TCNQ thin films grown on Au/Si(100) and on Si(100) substrates are compared with the reference data for pure TMP[20] and TCNQ,[21] simulated using the Mercury 2.2 program code.[22] $\theta$ is the angle between the incident ray and the scattering plane. A new crystallographic phase is evident, because new reflections corresponding to lattice plane spacings of $d_1$ = 0.894 nm and $d_2$ = 0.677 nm occur, as marked by the arrows. The diffractograms of the TMP-TCNQ thin films grown on Au/Si (topmost curve, thin film) and on Si (second curve, thicker film) reveal no pure TCNQ and only a very small amount of pure TMP coexisting with the new crystallographic phase. A detailed thin film growth analysis and structure resolution is presently in progress and will be presented in a forthcoming publication.

The structure resolution of the new CT compound is impeded by the presence of pure donor and acceptor phase in the grown TMP-TCNQ thin films. Therefore to get the crystal structure of the TMP-TCNQ complex growth optimization is needed. Presently, we can speculate on the likely stacking geometry of the newly found CT complex. In analogy to another CT compound containing the pyrene molecule: (hexamethoxypyrene - TCNQ)[20] a mixed stack geometry in TMP-TCNQ was found, but X-ray structure analysis needs futher refinement.

## 3. Spectroscopy of the pure and mixed phases
### 3.1 Infrared spectroscopy

Since a CT in the complex acts on the strength of chemical bonds, IR spectroscopy is a common tool to study this phenomenon in terms of the characteristic vibrational frequencies. For TCNQ, the CN stretching vibration has proven to be a useful indicator of a CT, see e.g. Ref. [8].

Infrared spectra have been taken using a Fourier spectrometer (type Nicolet 730 FT-IR) in reflection geometry. Fig. 2 shows the infrared spectra in the region of the CN stretching frequency for two different thin-film samples of the TCNQ-TMP mixed phase in comparison with pure TCNQ on Au. The spectrum of the low-coverage sample of the mixed phase (i) reveals a CN stretching frequency of 2219 cm$^{-1}$, being red shifted by 7 cm$^{-1}$ in comparison to pure TCNQ on Au (2226 cm$^{-1}$). This shift is characteristic for a CT to the acceptor TCNQ of the order of $0.3\,e$, as discussed in Ref. [8]. The spectrum of the high-coverage sample (ii) shows a coexistence of red-shifted peak at 2219 cm$^{-1}$ with the unshifted CN vibration (shoulder at 2226 cm$^{-1}$). The intensity ratio indicates that this sample is predominantly the TCNQ-TMP mixed phase, but contains an admixture of pure TCNQ. The spectrum of the low-coverage sample (i) reveals that this sample contains no admixtures. Such samples are produced at sufficiently low growth rates.

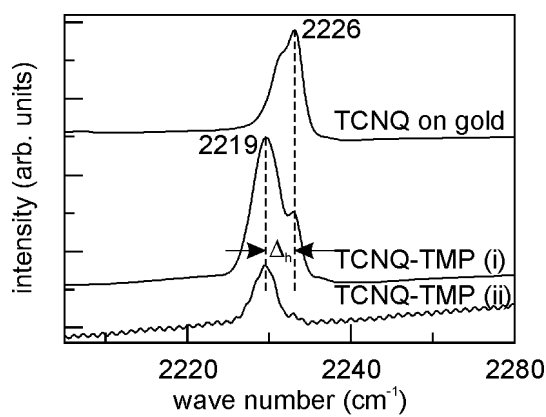

**FIG. 2.** IR spectra of the CN stretching vibration in a TCNQ film on Au and films of the mixed phase TCNQ$_1$-TMP$_1$ on Au at low coverage (i) and high coverage (ii). In the mixed phase the vibration frequency is red-shifted by $\Delta = 7$ cm$^{-1}$, being indicative of a CT of the order of $0.3\,e$.

The periodic oscillations superimposed to the spectrum of sample (i) originate from the underlying Au film. Given the photon impact angle normal to the surface and the periodicity of the interference fringes (2.5 cm$^{-1}$) we estimate a thickness of the underlying Au film of 15 nm. The low-coverage sample (i) contains crystallites of the mixed phase on top of a smooth multilayer (Fig. 1b, upper image), whereas in sample (ii) the surface is completely covered by crystallites (Fig. 1b, lower image).

*3.2 Ultraviolet photoelectron spectroscopy*

UPS measurements have been performed in a μ-metal-shielded UHV chamber equipped with a photoelectron spectrometer (FOCUS CSA 300) and a helium discharge lamp operated at the He I line ($h\nu = 21.23$ eV). Samples are introduced via a UHV load-lock. A *xyz*-manipulator with polar rotation (VG Omniax) allows in situ transfer from the prepchamber (where the Au films are deposited) to the spectrometer chamber (where the organic films are deposited in the measuring position). The photon beam is directed at grazing incidence (85° with respect to the surface normal of the sample). All data shown were taken under normal emission in order to obtain the correct values of the sample work function. The electric field vector of the radiation thus has a large component along the surface normal. The analyzer pass energy was set to 4 eV with 4 mm entrance and exit slits, yielding an energy resolution of < 0.1 eV and an angular resolution of ±3°. Further details on the experimental set-up are given in Ref. [5].

Fig. 3 shows series of photoelectron spectra taken during deposition of TMP (a) and TCNQ (b) on clean Au surfaces in comparison with theoretical spectra (top). The insets show the corresponding coverage-dependences of the work function $\Phi$, the numbers correspond to the spectra numbers.

For **TMP on Au** (Fig. 3a) the work function drops rapidly from 5.3 to 4.7 eV during deposition of the first monolayer. This constitutes the "standard" behaviour for polycyclic aromatic hydrocarbon molecules on noble metal surfaces. The drop in $\Phi$ reflects the push-back effect, i.e. Pauli repulsion of the spill-out charge of the metal surface by the valence electrons in molecules of the first monolayer. This reduces the surface dipole of the metal. The sequence of UPS spectra shows the growth of signals A, B and C at binding energies of 2.0, 3.0 and 4.6 eV below $E_F$ that do not change their positions as function of coverage.

The system **TCNQ on Au** (Fig. 3b) shows a more complex behaviour: During the initial phase of adsorption the work function $\Phi$ drops slightly by 0.1 eV and then shows a steep rise by almost 0.7 eV. The initial weak variation reflects the interplay between the push-back effect (lowering $\Phi$), a possible CT from Au into the first molecular layer (rising $\Phi$) and, in addition, a possible deformation of the molecule. Bent TCNQ on noble-metal surfaces has been described in Ref. [23]. Upon completion of the first monolayer the push-back effect has reached saturation because it is only caused by molecules in direct contact with the metal. However, the CT keeps on so that the work function exhibits a steep rise until the second layer is completed. Above two layers the further increase is rather small and the work function saturates at about 6.0 eV. This high value proves that there is a considerable CT from the metal into the organic film, leading to a strong dipole layer at the organic-metal interface. With increasing coverage the signals A and B grow at binding energies of 3.3 and 4.8 eV below $E_F$. Their positions do not change as function of coverage.

An important quantity for the electronic properties of organic films is the *hole injection barrier* $\Delta_h$, defined as the distance between the leading edge of the HOMO and the Fermi-energy of the metal. From Fig. 3 we obtain $\Delta_h$ = 1.6 and 2.7 eV for TMP and TCNQ, respectively. Together with the measured work functions (see insets) we derive ionization energies of $E_i = \Delta_h + \Phi =$ 6.3 and 8.7 eV for TMP and TCNQ molecules in multilayer films, respectively.

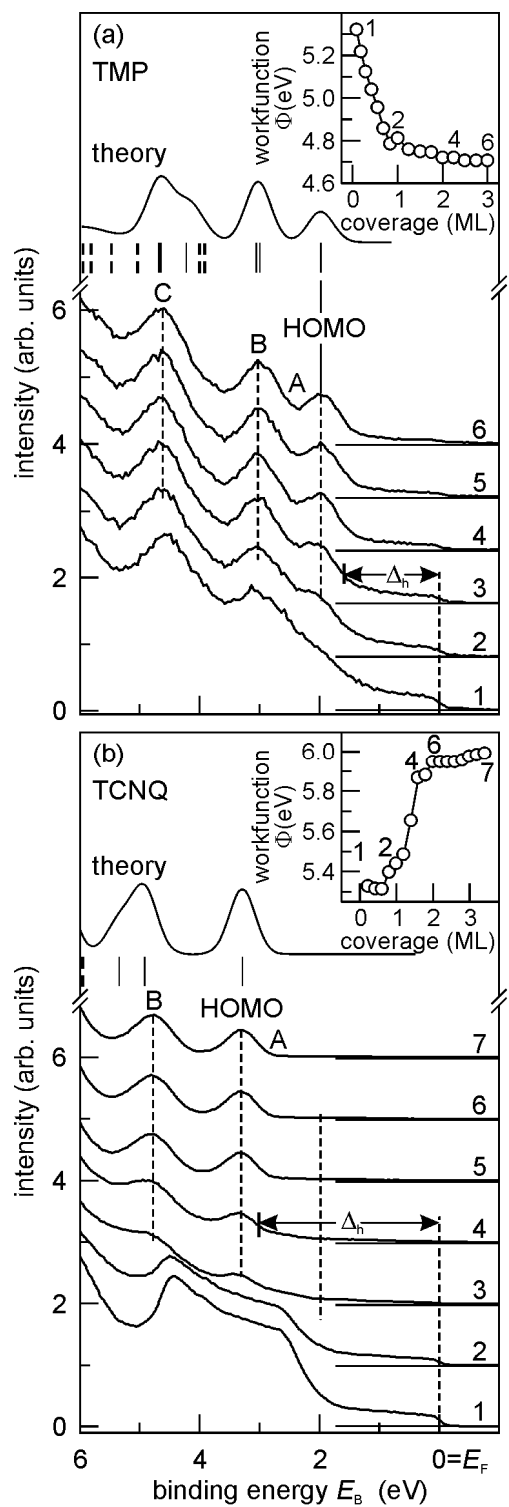

**FIG. 3.** Coverage series of He I UPS spectra of TMP (a) and TCNQ on Au (b). For comparison, spectra calculated on the basis of ΔSCF binding energies are included (top) that agree well with the measured spectra. In order to align the HOMO, the theoretical spectra have been rigidly shifted by 0.52 and 0.48 eV to lower binding energy for TMP and TCNQ, respectively. Full and dotted bars denote the binding energy positions of π- and σ-like orbitals, respectively. The insets show the corresponding workfunction-vs-coverage curves (coverage in monolayers ML), numbers correspond to the spectra numbers. $\Delta_h$ denotes the hole injection barriers.

Theoretical binding energies $E_B$ of electrons in the molecular orbitals were calculated in Delta Self Consistent Field (ΔSCF) approximation with explicit computation of energies of both the neutral compound and resulting cation after the vertical photo-transition, using the B3LYP/6-31G(d) correlation functional (details are given in Ref. [24]). The electrons in the HOMO have theoretical binding energies of $E_B$ = 6.18 and 8.87 eV for TMP and TCNQ, respectively. Taking into account the work functions of the multilayer films ($\Phi_{TMP}$ = 4.7 eV and $\Phi_{TCNQ}$ = 6.0 eV) these values translate into binding energies of 1.48 and 2.87 eV with respect to the Fermi-energy. The calculated binding energies are 0.52 eV (for TMP) and 0.48 eV (for TCNQ) lower than the maxima positions of the measured HOMO signals (peaks A).

The theoretical spectra have thus been aligned at the experimental HOMO positions, i.e. they are rigidly shifted by about half an eV to higher binding energies. The calculated spectra assume a Gaussian broadening of 0.5 eV FWHM. The partial cross sections of the orbitals with π-like symmetry (full bars) were taken to be identical; in particular no degeneracy exists for these molecules. Actually there is no strict π- and σ-symmetry in TMP because of the methyl groups. However, we can identify "π- and σ-like" symmetry by analogy with the parent molecule pyrene. The oxygen 2p-derived orbitals with σ-like symmetry (dashed bars in Fig. 3a) deserve special consideration. At the given photon energy the partial cross section of oxygen 2p is lower than that of carbon π.[25] Moreover, the σ-orbitals have nodal planes perpendicular to the molecular plane. Hence the corresponding photoemission signals in normal emission vanish for molecules oriented parallel to the surface. In order to account for the lower O 2p cross section and the orientational effect, the corresponding signal intensities were weighted by a factor of 0.2 in comparison with the π-orbitals. The orbital at 5.3 eV in TCNQ has a high contribution from nitrogen 2p. Likewise, this cross section is lower and the spectral intensity of the 5.3 eV line has thus been weighted by a factor of 0.5.

Except for the shifts aligning the HOMO levels with the experimental signals A, both calculated spectra agree well with the measured spectra. Signal B results from the closely spaced HOMO-1/HOMO-2 orbitals and signal C represents the group of HOMO-3 to HOMO-7. For TCNQ signal B corresponds to HOMO-1 and the nitrogen 2p-derived contribution of HOMO-2 in the shoulder.

This shows that the ΔSCF calculation is well suited to analyze the frontier orbital structure of these molecules. The essential reason for the difference between absolute level positions in theory and experiment is the fact that the calculation assumes a single free molecule, whereas the measurements are made for a condensed film on a metal surface. It is interesting to note that the same theory shows a significantly better quantitative agreement for the larger PAHs hexamethoxy- and hexaketocoronene in condensed films.[5,23]

Fig. 4 shows a sequence of photoelectron spectra for the **mixed phase of TMP-TCNQ** in the range from sub-monolayer to multilayer coverage. The inset shows the corresponding work function-vs-coverage curve, with numbers corresponding to the numbers at the spectra. Adsorption in the low-coverage regime is characterized by the growth of signals A´ and B´ at 1.5 and 2.5 eV below $E_F$ and by an increase of the workfunction by 0.4 eV. Obviously, the organic-metal interface of the mixed phase is polar due to a CT from Au to the overlayer. The Fermi edge vanishes rapidly with increasing coverage, pointing on *complete wetting*. In spectrum 5 this low-coverage phase is completed, i.e. signals A´ and B´ have reached their intensity maxima and the rise of Φ abruptly ends. The hole injection barrier in this growth regime is $\Delta_h = 1.0$ eV.

The higher coverage regime is characterized by a weak further increase of Φ and by the growth of signals A and B, being markedly shifted with respect to the low-coverage signals A´ and B´. Φ reaches a saturation value of 5.75 eV and signals A and B are fully developed at 1.8 and 2.9 eV below $E_F$ (spectrum 8). We attribute the energy level shift between the low-coverage and higher-coverage regime to a change in the film morphology. In addition, the increased image charge screening of the photo-hole in the first layer (typical for adsorbates on metal surfaces)[26] might contribute to the shift. Spectrum 6 shows the coexistence of peaks A/B and A´/B´. This spectrum corresponds to a structure similar to sample (i) in Fig. 1b that exhibits both smooth areas and 3D crystallites. We thus conclude that signals A´ and B´ result from the initial smooth multilayer, whereas A and B in spectra 7-9 originate from a surface completely covered by three-dimensional TCNQ-TMP crystallites (cf. sample (ii) in Fig.1b). The development of UPS spectra in Fig. 4 and the AFM images in Fig. 1b suggest a Stranski-Krastanov like growth mode for the mixed phase of TCNQ-TMP. A similar growth mode has been observed for hexabenzocoronene on Au (111), where in the first four layers the molecules lie flat on the surface, whereas above four layers 3D bulk crystallites develop.[27] Previous STM investigations showed that the mixed phases of $F_4$TCNQ and α-sexithiophene[18] as well as TCNQ and TTF[19] adsorb in smooth monolayers, where the donor and acceptor molecules form regular patterns.

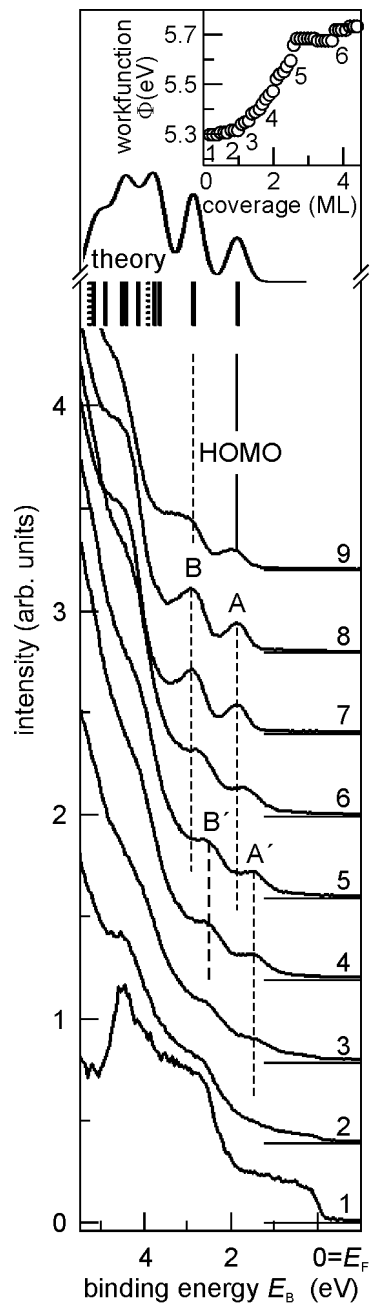

**FIG. 4.** Coverage series of UPS spectra taken during the growth of the TCNQ-TMP mixed phase on Au. The inset shows the corresponding workfunction curve, numbers correspond to the spectra numbers. For comparison, a theoretical spectrum is shown (top), calculated on the basis of ΔSCF binding energies. In order to align the HOMO with the experimental signal A the theoretical spectrum has been rigidly shifted by 0.97 eV to higher binding energy.

Spectrum 9 shows a drop in intensity of signals A and B and a broadening of signal B towards higher binding energies (wing to the left). One reason could be the formation of coexisting crystallites of TCNQ, in accordance with the IR measurement, see spectrum (ii) in Fig. 2. The calculation reveals that A constitutes the HOMO, whereas B is composed of HOMO-1 and HOMO-2, being closely spaced in energy. The theoretical spectrum in Fig. 4 has been rigidly shifted by 0.97 eV to lower binding energies, in order to align the HOMO position with peak A. Like for the pure species, no perfect agreement can be expected because the calculation assumes an isolated TCNQ-TMP complex molecule instead of a 3D crystal. Energy minimization yields a parallel orientation with the centres of donor and acceptor lying above each other with 0.363 nm spacing and 0.6 nm lateral displacement.

### *3.3 Scanning tunnelling spectroscopy*

For STS we deposited TMP-TCNQ in UHV under the same conditions as for the previously described experiments. In this case coverages of TMP-TCNQ corresponding to the coverage regime (i) in Fig. 1b were deposited on a clean W(110) surface for which a similar substrate - adsorbate interaction is assumed as in the case of the Au(111) surface used before. In particular, the work functions are almost identical (5.3 and 5.2 eV for the Au films and W(110), respectively). STS was performed to obtain differential conductance $dI/dU$ maps and spectra, using a lock-in technique with a 8 kHz bias voltage modulation of 30-50 mV. All bias voltages given are sample voltages with respect to the tip. We used a $Pt_{80}Ir_{20}$ tip that was cut under tensile stress from a thin wire. Spectroscopic data shown here were measured at room temperature directly after the sample preparation with residual gas exposures less than 0.5 L (1 L= $10^{-6}$ Torr s). Spectroscopic $dI/dU$ curves were recorded on top of homogeneous topographically elevated structures which were identified as areas covered by TMP-TCNQ. The stabilization parameters ($U$ = 0.7 V, $I$ = 0.35 nA) and modulation amplitude (50 mV) for the results shown here were kept constant. In order to reduce the noise we averaged spectra taken at random positions. Fig. 5 shows the results for the TMP-TCNQ samples. While the spectra were reproducible for sample voltages -1 V < $U$ < 3 V, the spectroscopic features observed for $U$ < -1 V appeared to be strongly influenced by tip induced states.

In a simple model, the tunnelling conductance is roughly proportional to the local density of states (DOS) of the sample at the tip position multiplied by the tunnelling probability. For positive (negative) sample bias one measures unoccupied (occupied) sample states assuming a flat DOS for the tip. Pronounced peaks are often observed for surface states or resonant states leaking out into vacuum.

We observe a low and flat differential conductivity between -0.4 and +0.2 eV indicating a low DOS. This region is confined by a pronounced peak at +0.50 eV and a broader peak with a maximum at ca. -0.75 eV. Both features are absent on the clean W(110) surface. Therefore, we identify these peaks as the LUMO and HOMO state of the organic overlayer. Due to the interaction of the molecular states with the metallic substrate one expects a shift of all molecule states relative to the situation for free molecules or to thicker coverages. However, to first approximation one expects the same shift for all states keeping the differences between the energies constant. The difference of the energies of the two peak maxima (1.25 eV) defines the energy gap between HOMO and LUMO for the CT compound. The hole-injection barrier determined by STS ($\Delta_h$ = 0.5 eV in Fig. 5) is significantly smaller than the value resulting from UPS ($\Delta_h$ = 1 eV in Fig. 4). The electron injection barrier, i.e. the distance between the metal Fermi edge and the onset of the LUMO is only $\Delta_e$ = 0.3 eV.

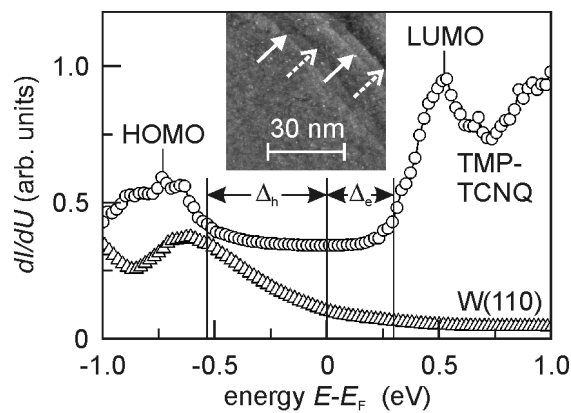

**FIG. 5.** Differential conductivity $dI/dU$ spectrum measured at room temperature on an island of TMP-TCNQ, compared with a spectrum from the clean W(110) substrate. Data shown result from an average of individual spectra measured on homogeneous topographically elevated areas of the sample (for clarity, the TMP-TNCQ spectrum was shifted upwards by 0.2 arb. units). The sample voltage modulation was 50 mV and the tip was stabilized at $U$ = 0.7 V and $I$ = 0.35 nA. The inset shows a processed topographical STM image of the corresponding region where the spectra were measured. Full (dashed) arrows indicate monoatomic substrate (molecular complex) steps indicating an initial step flow growth mode. $\Delta_h$ and $\Delta_e$ denote the hole- and electron-injection barriers, respectively.

*3.4 Comparison of experimental and theoretical level positions*

Fig. 6 summarizes the measured and calculated level positions for the acceptor molecule TCNQ (left column), the donor molecule TMP (right column) and their CT complex (centre column). In order to be able to compare experiment and theory the energy scale is referenced to the vacuum level. In Figs. 3-5 the UPS and STS data are referenced to the Fermi level of the gold substrate. A change of the surface dipole induced by the organic film leads to an according shift of the energy levels that are pinned to the new electrostatic potential in the organic film, as discussed in Ref. [16]. The energy levels of the molecules are "biased" by the surface dipole. For TCNQ the CT from the metal into the first layers of the film leads to an interface dipole and the work function increases. For TMP the push-back effect (Pauli repulsion) acting on the spill-out electrons in front of the surface by the molecular electrons leads to a reduction of the metal surface dipole so that the work function decreases. The change of the interface dipole cannot be modelled by theory. Therefore, these work function changes are not taken into account in Fig. 6 for sake of a quantitative comparison with theory and the experimental peak energies are given with respect to $E_F(Au)= 5.3$ eV.

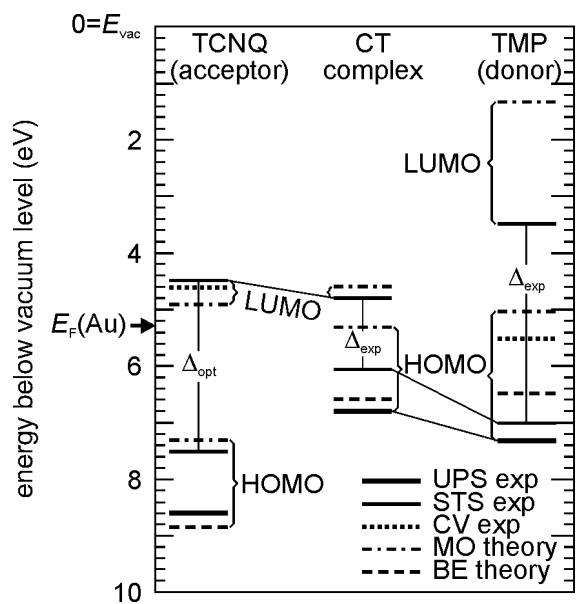

**FIG. 6.** Energy scheme of the TCNQ-TMP complex (centre) in comparison with the pure phases of TCNQ (left) and TMP (right). In order to compare with theory, the binding energy scale is referenced to the vacuum level. Thick, thin and dotted lines denote the experimental UPS, STS and CV results, chain and dashed lines represent theoretical molecular orbital eigenvalues and $\Delta$SCF binding energies, respectively. $\Delta_{opt}$ is the experimental value[20] for the optical gap of TCNQ, adapted at the STS LUMO level.[29] $\Delta_{exp}$ denote the gaps determined by STS for TMP and the complex.

For the *acceptor TCNQ* the LUMO determined by the STS-measurement[28] (full line) and the CV result[29] (dotted line) agree very well and lie close to the prediction of the molecular orbital calculation (chain line). The HOMO position observed in UPS (thick line) agrees quite well with the calculated $\Delta$SCF binding energy (dashed). The measured optical gap[30] ($\Delta_{opt}$ = 3.2 eV) is significantly larger than the difference of the HOMO-LUMO eigenvalues (chain lines, distance 2.4 eV). Due to its large distance from the Fermi level, the HOMO of TCNQ could not be observed in STS. The HOMO position denoted by the thin line at 7.5 eV was derived from the LUMO position measured by STS, adding the value of the optical gap $\Delta_{opt}$.

For the *donor TMP* the calculated level positions deviate considerably from the measured results. The LUMO (HOMO) eigenvalues lie 2.5 eV (2 eV) higher than the measured STS values. The measured UPS HOMO position lies rather close to the STS value (0.3 eV difference).

The calculated ΔSCF binding energy is about 0.6 eV smaller than the STS/UPS signals. The CV result for the HOMO lies 1.8 eV above the STS/UPS signals, but quite close to the calculated HOMO eigenvalue. Obviously there is a substantial difference between the measurements for the condensed phase (STS, UPS) of TMP on the one hand and the calculation for free molecules and the CV measurement in liquid environment on the other hand. The size of the gap as determined by STS ($\Delta_{exp}$ = 3.5 eV) compares well with the measured optical gap[20] of 3.16 eV and agrees well with the difference of the HOMO-LUMO eigenvalues (3.6 eV).

For the *CT complex* STS reveals a strong reduction of the gap to $\Delta_{exp}$ = 1.25 eV due to an upward shift of the energy of the HOMO (compared with TMP) by 1.0 eV and a smaller downward shift of the LUMO (compared with TCNQ) by 0.3 eV. The calculation shows that the HOMO (LUMO) of the complex is derived from the respective orbitals of TMP (TCNQ). The upward shift of the HOMO is also visible in UPS (thick lines). Since STS values have been taken at low coverage, Fig. 6 shows the low-coverage UPS signal A´ from Fig. 4.

For the complex the difference between the STS and UPS HOMO positions (distance 0.6 eV) is somewhat larger than for TMP. The ΔSCF binding energy prediction (dashed) lies close to the UPS signal. The calculated LUMO eigenvalue (upper chain line) shows good agreement with the STS LUMO position, similar as in the case of TCNQ. The HOMO eigenvalue is substantially too high, i.e. the calculated gap of 0.6 eV is about a factor of 2 smaller than the experimental gap ($\Delta_{exp}$ = 1.25 eV). The HOMO of TCNQ becomes HOMO-3 in the complex; it is not shown in Fig. 6.

## 4. Summary and Conclusion

We have studied UHV co-deposited films of the novel donor TMP and the classical acceptor TCNQ in 1:1 stoichiometry on Au and Si using X-ray diffraction and various spectroscopic techniques. New reflexes corresponding to periodicities of $d_1$ = 0.894 nm and $d_2$ = 0.677 nm in X-ray diffraction (θ - 2θ scans) revealed a new crystallographic phase. Infrared spectroscopy shows a softening of the CN stretching vibration of TCNQ by 7 cm$^{-1}$ in the complex as compared with pure TCNQ on Au. This shift is indicative of a CT of the order of $0.3\,e$. The occupied and unoccupied frontier orbitals of the CT complex and the pure moieties have been probed by UPS and scanning tunnelling spectroscopy STS, revealing characteristic level shifts and a strong reduction of the HOMO-LUMO gap upon formation of the complex.

In order to interpret the spectra, DFT calculations have been performed using the Gaussian03 code with the B3LYP hybrid functional. The comparison of measured and calculated energy levels of the CT complex and the pure donor and acceptor shows the following: The complex is characterised by a relatively small HOMO-LUMO gap of 1.25eV (referred to the positions of the peak maxima). The calculated LUMO eigenvalue lies close to the LUMO peak measured by STM, the HOMO eigenvalue lies 0.7 eV above the STM HOMO peak; the calculated HOMO-LUMO gap is about a factor of 2 too small. The calculated ΔSCF binding energy of the electrons in the HOMO agrees fairly well with the UPS peak position. Taking the hole-injection and electron-injection barriers $\Delta_h$ and $\Delta_e$ from the STS spectrum in Fig. 5, we estimate an onset of the transport gap of about $\Delta_{transp} \cong \Delta_h + \Delta_e =$ 0.8 eV, making TCNQ-TMP a small-band-gap semiconductor. Measurements and theory clearly show that the small gap arises from the fact that the TMP-derived HOMO shifts towards lower binding energy and the TCNQ-derived LUMO shifts to higher binding energy as compared with the pure moieties. The given values correspond to the low-coverage regime. At higher coverage the HOMO shifts by about 0.4 eV to higher binding energy (A, B in Fig. 4), so that the gap will be somewhat larger for thick films or for bulk material of the CT compound. Pure TCNQ and TMP exhibit much larger gaps of more than 3 eV.

For TCNQ, we find good agreement between calculated and measured level positions: The STS LUMO lies close to the calculated LUMO eigenvalue and almost coincides with the CV result. Like for the complex, the UPS HOMO peak lies quite close to the predicted ΔSCF binding energy. The calculated HOMO-LUMO gap is only 20% smaller than the experimental value for the optical gap from Ref. [30].

For TMP the deviations between experiment and theory are larger. The HOMO and LUMO eigenvalues lie more than 2 eV above the STS values and the ΔSCF binding energy of the HOMO lies 0.8 eV above the UPS signal. However, the CV value for the HOMO is fairly close to the HOMO eigenvalue. We may speculate that the pinning-effect of the energy levels in the thin film to the new electrostatic potential in the organic film due to the organic-metal interface (as discussed in Ref. [16]), might be responsible for this discrepancy. This pinning is not accounted for by theory and CV measurements are free of such interface effects. This is probably the main reason why the experimental data for the condensed film lie significantly below the calculated values and the CV result.

Besides these near-at-hand differences between condensed and free molecules, there are also *systematic differences between the STS and UPS positions of the HOMOs*. The HOMO levels measured by STS for TMP and for the complex as well as the HOMO position for TCNQ estimated using $\Delta_{opt}$ (thin line) lie consistently above the UPS results, the distance being 0.7, 0.6 and 0.3 eV for TCNQ, the CT-complex and TMP, respectively. The differences point on the fact that these two spectroscopies are fundamentally different: STS probes the *ground-state HOMO position via resonant tunnelling from the HOMO into the tip*. The tunnelling matrix element is governed by the overlap integral of (ground-state) wavefunctions in the tip and the molecular film. UPS yields the energies of the *final (ionic) states being screened by neighbouring molecules*. The electronic transitions in UPS are driven by the photon operator, in the low energy regime of UPS the dipole operator *er* is a very good approximation. Energetically, UPS measures the energy difference between the initial neutral $N$ electron state and the final $N$-1 electron state. Although both spectroscopies probe "the HOMO", the energy positions are expected to be quantitatively different. E.g. the polarization energy,[31] being a significant contribution in UPS from condensed layers (in comparison with free molecules) is absent for STS. The long-range Coulomb attraction acting on the outgoing photoelectrons reduces their kinetic energy. This translates into higher UPS binding energies in Fig. 6, in accordance with the observed systematic differences between UPS (thick lines) and STS (thin lines).

The fact that the deviation between UPS and STS HOMO positions is maximum for TCNQ might be connected with the strong interface dipole increasing the work function by 0.7 eV. In this case, the negative side of the dipole is oriented towards the organic film. It is likely that STS experiences only part of this "level biasing" because the tip itself introduces an energy reference level towards the vacuum side, whereas this reference level is missing in UPS.

For the CT-compound we observed the growth of a flat multilayer at low coverage and the appearance of 3D crystallites above about 4 monolayers. It is likely that the molecules in the first layers lie in a face-on orientation on the surface, whereas the multilayer film attains the bulk structure of the compound. This structural transition from the smooth multilayer to 3D crystallite growth shows up in UPS as a shift of the photoemission peaks towards higher binding energy. In this context it is worthwhile being mentioned that the positions of peak maxima and line shapes in angular resolved UPS of thin films depend on the molecular orientation in the film.[32-34]. This effect can also be interpreted in terms of orientation-dependent ionization energies.[35] A full structural analysis of the studied thin films is thus highly desirable, such a study is presently being performed.

**Acknowledgements**

We thank Norbert Koch (Humboldt Universität Berlin) for fruitful discussions. The project is funded through Transregio SFB TR 49 (Frankfurt, Mainz, Kaiserslautern), Graduate School of Excellence MAINZ and Centre for Complex Materials (COMATT), Mainz.

---

\* FAX: +49-6131-3923807; Email address: schoenhe@uni-mainz.de